\documentclass[conference, 9pt]{IEEEtran}
\IEEEoverridecommandlockouts
\usepackage{cite}
\usepackage{amsmath,amssymb,amsfonts}
\usepackage{algorithmic}
\usepackage{graphicx}
\usepackage{textcomp}    
\usepackage{xcolor}
\usepackage{enumitem}
\usepackage{multirow}
\def\BibTeX{{\rm B\kern-.05em{\sc i\kern-.025em b}\kern-.08em
    T\kern-.1667em\lower.7ex\hbox{E}\kern-.125emX}}
\usepackage{tikz}

\newcommand*\circledB[1]{\tikz[baseline=(char.base)]{
            \node[shape=circle,fill,inner sep=0.2pt] (char) {\textcolor{white}{#1}};}}

\usepackage{booktabs}

\usepackage{fancyhdr}
\fancypagestyle{firstpage}{
	\fancyhf{}
	\chead{\small To appear at the 40th IEEE/ACM International Conference on Computer-Aided Design (ICCAD), November 2021, Virtual Event.}
	\cfoot{\thepage}
}
 
\begin{document}

\title{Towards Energy-Efficient and Secure Edge AI:\\
A Cross-Layer Framework\\
\vspace{-2pt}
{\large ICCAD Special Session Paper}
\vspace{-8pt}
}

\author{\IEEEauthorblockN{Muhammad Shafique\IEEEauthorrefmark{1}, Alberto Marchisio\IEEEauthorrefmark{2}, Rachmad Vidya Wicaksana Putra\IEEEauthorrefmark{2}, Muhammad Abdullah Hanif\IEEEauthorrefmark{2}\IEEEauthorrefmark{1}}
\IEEEauthorblockA{\IEEEauthorrefmark{1}\textit{New York University Abu Dhabi (NYUAD)},
Abu Dhabi, United Arab Emirates}
\IEEEauthorblockA{\IEEEauthorrefmark{2}\textit{Technische Universit{\"a}t Wien (TU Wien)},
Vienna, Austria}
muhammad.shafique@nyu.edu, \{alberto.marchisio, rachmad.putra\}@tuwien.ac.at, mh6117@nyu.edu %
\vspace{-0.3cm}
}

\maketitle
\thispagestyle{firstpage}

\begin{abstract}
The security and privacy concerns along with the amount of data that is required to be processed on regular basis has pushed processing to the edge of the computing systems. 
Deploying advanced Neural Networks (NN), such as deep neural networks (DNNs) and spiking neural networks (SNNs), that offer state-of-the-art results on resource-constrained edge devices is challenging due to the stringent memory and power/energy constraints. 
Moreover, these systems are required to maintain correct functionality under diverse security and reliability threats. 
This paper first discusses existing approaches to address energy efficiency, reliability, and security issues at different system layers, i.e., hardware (HW) and software (SW). 
Afterward, we discuss how to further improve the performance (latency) and the energy efficiency of Edge AI systems through HW/SW-level optimizations, such as pruning, quantization, and approximation. 
To address reliability threats (like permanent and transient faults), we highlight cost-effective mitigation techniques, like fault-aware training and mapping. 
Moreover, we briefly discuss effective detection and protection techniques to address security threats (like model and data corruption). 
Towards the end, we discuss how these techniques can be combined in an integrated cross-layer framework for realizing robust and energy-efficient Edge AI systems.
\end{abstract}

\begin{IEEEkeywords}
Artificial intelligence, machine learning, Edge AI, deep neural networks, spiking neural networks, accuracy, latency, energy efficiency, reliability, security, robustness, Edge computing, tinyML.
\end{IEEEkeywords}

\section{Introduction}

The NN-based AI systems have achieved state-of-the-art accuracy for various applications such as image classification, object recognition, healthcare, automotive, and robotics~\cite{Ref_LeCun_DL_Nature15}.
However, current trends show that the accuracy is improved at the cost of increasing complexity of NN models (e.g., larger model size and complex operations)~\cite{Ref_Capra_DNNSurvey_Access20}\cite{Ref_Putra_FSpiNN_TCAD20}. 
This increased complexity hinders the deployment of advanced NNs (DNNs and SNNs) on resource-constrained edge devices~\cite{Ref_Marchisio_DLforEdge_ISVLSI19}.
Therefore, optimizations at different system layers (i.e., HW and SW) are necessary to enable the use of advanced NNs at the edge~\cite{Ref_Capra_DNNSurvey_Access20}.
Besides performance and energy efficiency, reliability and security aspects are also important to ensure the correct functionality under diverse operating conditions (e.g., in the presence of HW faults and security threats), especially for safety-critical applications like autonomous driving and healthcare~\cite{Ref_Kriebel_RobustnessML_ISVLSI18}\cite{Ref_Shafique_RobustML_DnT20}.
Therefore, \textit{the important design metrics for enabling Edge AI include performance (i.e., latency), energy efficiency, reliability, and security}. 

\subsection{Key Challenges for Energy-Efficient and Secure Edge AI}

\begin{figure}[t]
\centering
\includegraphics[width=\linewidth]{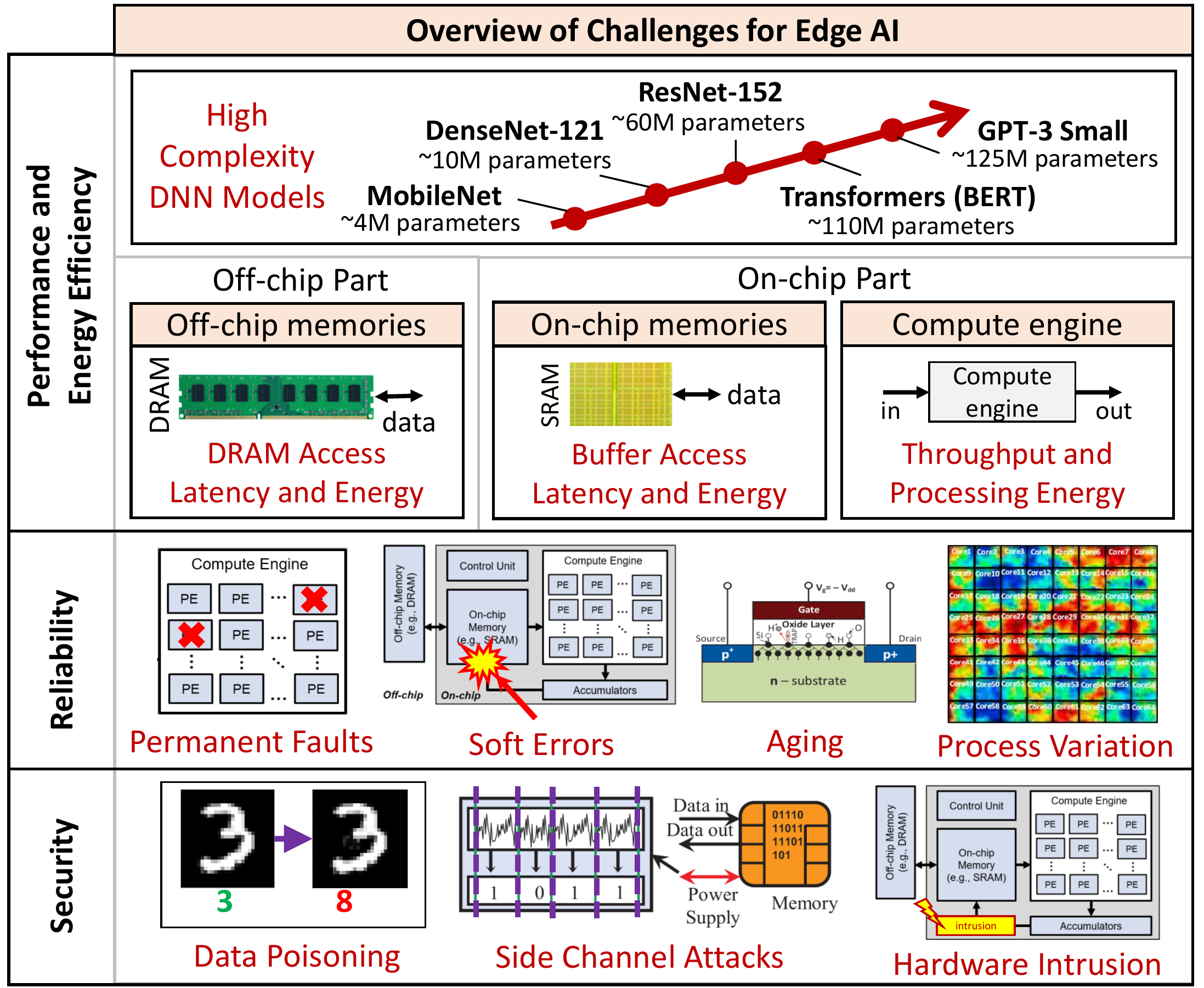}
\vspace{-0.5cm}
\caption{Overview of challenges for energy-efficient and secure Edge AI.}
\label{Fig_Challenges_Overview}
\vspace{-0.6cm}
\end{figure}

We introduce the key challenges for developing Edge AI systems in the following text (see Fig.~\ref{Fig_Challenges_Overview} for an overview of the challenges).   
  
\begin{itemize}[leftmargin=*]
    \item \textbf{Performance:} 
    Edge AI systems are expected to have high performance to provide real-time response.  
    However, due to memory- and compute-intensive nature of NNs, achieving high performance is not trivial. 
    Moreover, edge devices have limited compute and memory resources, which makes it challenging to map the full NN computations simultaneously on an accelerator fabric \cite{Ref_Putra_ROMANet_TVLSI21}.
    \item \textbf{Energy Efficiency:} 
    Edge AI systems should also have high energy efficiency to ensure complete processing within a restricted energy budget, especially in the case of battery-powered devices. 
    Therefore, the energy consumption in both the off-chip and on-chip parts should be minimized. 
    The off-chip part includes the DRAM-based off-chip memory accesses~\cite{Ref_Putra_ROMANet_TVLSI21}, while the on-chip part includes (1) the on-chip memory accesses, and (2) the neural operations like multiply-and-accumulation (MAC)~\cite{Ref_Marchisio_DESCNet_TCAD20}. 
    \item \textbf{Reliability:} 
    Edge AI systems should produce correct outputs even in the presence of different types of reliability threats~\cite{Ref_Hanif_RobustML_IOLTS18}. The main reliability threats are as follows.
    \begin{itemize}
        \item \textit{\textbf{Process variations}} are the result of imprecisions in the fabrication process, as manufacturing billions of nano-scale transistors with identical electrical properties is difficult  to impossible. 
        This causes variations in the leakage power and frequency in the same chip, across different chips in the same wafer, and even across different wafers~\cite{Ref_Raghunathan_CherryPicking_DATE13}.
        \item \textit{\textbf{Soft errors}} are caused by high-energy particle strikes, manifest as bit-flips at the HW layer, and can propagate all the way to the application layer and may cause incorrect outputs~\cite{Ref_Baumann_SoftErrors_TDMR05}.
        \item \textit{\textbf{Aging}} is gradual degradation of the processing circuits over time~\cite{Ref_Shafique_EDAchallenges_DAC14}. It occurs due to physical phenomena like Hot Career Injection (HCI), Time-Dependent Dielectric Breakdown (TDDB), and Negative/Positive Bias Temperature Instability (NBTI/PBTI). 
    \end{itemize}
    \item \textbf{Security:} 
    Edge AI systems should offer high resilience against security vulnerabilities such as side channels and HW intrusions~\cite{Ref_Hanif_RobustML_IOLTS18}. 
    Moreover, NN algorithms (e.g., DNNs) have other security vulnerabilities as well that can be exploited through data poisoning to cause confidence reduction or misclassification~\cite{Ref_Shafique_NextGenML_DATE18}. 
\end{itemize}


The above discussion highlights different possible challenges for developing energy-efficient and secure Edge AI systems. 
To address each challenge individually, various techniques have been proposed at different layers of the computing stack. \textit{However, systematic integration of the most effective techniques from both the hardware and software levels is important to achieve ultra-efficient and secure Edge AI.}

\subsection{Our Contributions}
In the light of the above discussion, the \textbf{contributions} of this paper are the following. 

\begin{itemize}[leftmargin=*]
    \item We present an overview of different challenges and state-of-the-art techniques for improving performance and energy efficiency of Edge AI systems~\textbf{(Section~\ref{Sec_PerformanceAndEnergy})}.
    \item We present an overview of different challenges and state-of-the-art techniques for reliability and security of Edge AI \textbf{(Section~\ref{Sec_ReliabilityAndSecurity})}.
    \item We present a cross-layer framework that systematically integrates the most effective techniques for improving the energy efficiency and robustness of Edge AI~\textbf{(Section~\ref{Sec_IntegratedFrameworks})}.
    \item We discuss the challenges and recent advances in neuromorphic computing considering SNNs~\textbf{(Section~\ref{Sec_NeuromorphicSNNs})}.
\end{itemize}

\renewcommand{\headrulewidth}{0pt}
\section{Performance and Energy-Efficient Edge AI}
\label{Sec_PerformanceAndEnergy}

\begin{figure}[t]
\centering
\includegraphics[width=\linewidth]{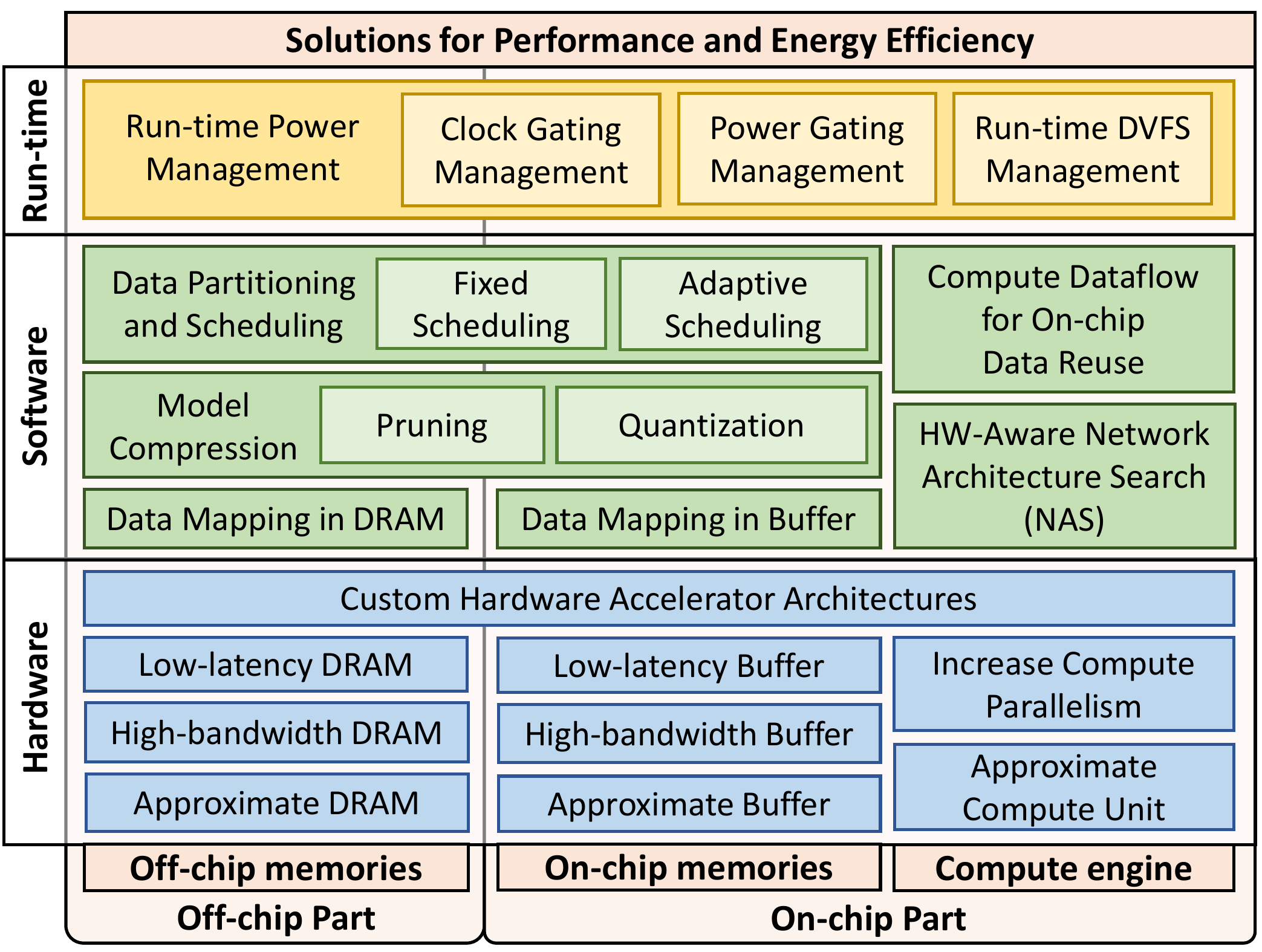}
\vspace{-0.6cm}
\caption{Overview of the techniques at different system layers for improving the performance and energy efficiency of Edge AI.}
\label{Fig_Challenges_PerformEnergy}
\vspace{-0.4cm}
\end{figure}

In the quest of achieving higher accuracy, the evolution of DNNs has seen a dramatic increase in the complexity with respect to model size and operations, i.e., from simple Multi-Layer Perceptron (MLP) to deep and complex networks like Convolutional Neural Networks (CNNs), Transformers, and Capsule Networks (CapsNet) \cite{Ref_Sabour_CapsNet_NIPS17}.
Hence, \textit{the advanced DNNs require specialized hardware accelerators and optimization frameworks to enable efficient and real-time data processing at the edge.}
To address this, a significant amount of work has been carried out in the literature. 
In this section, we discuss different state-of-the-art techniques for improving performance and energy efficiency of Edge AI (see overview in Fig.~\ref{Fig_Challenges_PerformEnergy}). 

\subsection{Optimizations for DNN Models}
\label{Sec_PerformanceAndEnergy_DNNmodel}

The edge platforms typically have limited memory and power/energy budgets, hence small-sized DNN models with limited number of operations are desired for Edge AI applications. 
Model compression techniques such as pruning (i.e., structured \cite{Ref_Anwar_StructuredPruning_JETC17}\cite{Ref_He_AMC_ECCV18} or unstructured \cite{Ref_Li_Pruning_arXiv16,Ref_Marchisio_PruNet_IJCNN18,Ref_Han_DeepCompression_ICLR15}) and quantization \cite{Ref_Han_DeepCompression_ICLR15,Ref_Gupta_LimitedPrecision_ICML15,Ref_Jacob_Quantize_CVPR18,Ref_Marchisio_Q-CapsNets_DAC20} are considered to be highly effective for reducing the memory footprint of the models as well as for reducing the number of computations required per inference. 
Structured pruning~\cite{Ref_Anwar_StructuredPruning_JETC17} can achieve about 4x weight memory compression, while class-blind unstructured pruning (i.e., PruNet~\cite{Ref_Marchisio_PruNet_IJCNN18}) achieves up to 190x memory compression. 
%
Quantization when combined with pruning can further improve the compression rate. 
For instance, quantization in the Deep Compression~\cite{Ref_Han_DeepCompression_ICLR15} improves the compression rate by about 3x for the AlexNet and the VGG-16 models. 
The Q-CapsNets framework~\cite{Ref_Marchisio_Q-CapsNets_DAC20} shows that quantization is highly effective for complex DNNs such as CapsNets as well. It reduces the memory requirement of the CapsNet~\cite{Ref_Sabour_CapsNet_NIPS17} by 6.2x with a negligible accuracy degradation of 0.15\%.
Since model compression techniques may result in a sub-optimal accuracy-efficiency trade-off (due to lack of information of the underlying hardware architecture used for DNN execution), HW-aware model generation and compression techniques have emerged as a potential solution. 
Many Neural Architecture Search (NAS) techniques~\cite{Ref_Zoph_NASNet_arXiv16,Ref_Liu_DARTS_arXiv18,Ref_Pham_ENAS_ICML18,Ref_Liu_PNAS_ECCV18,Ref_Tan_MnasNet_CVPR19, Ref_Achararit_APNAS_Access20,Ref_Marchisio_NASCaps_ICCAD20} have been proposed to generate high accuracy and efficient models. 
The state-of-the-art NAS like the APNAS framework~\cite{Ref_Achararit_APNAS_Access20} employs an analytical model and a reinforcement learning engine to quickly find DNNs with good accuracy-efficiency trade-offs for the targeted systolic array-based HW accelerators~\cite{Ref_Hanif_MPNA_arXiv18}\cite{Ref_Jouppi_TPU_ISCA17}.
It reduces the compute cycles by 53\% on average with negligible accuracy degradation (avg. 3\%) compared to the state-of-the-art techniques.
Therefore, it is suitable for generating DNNs for resource-constrained applications.
Meanwhile, the NASCaps framework~\cite{Ref_Marchisio_NASCaps_ICCAD20} employs an NSGA-II algorithm to find  Pareto-optimal DNN models by leveraging the trade-off between different hardware characteristics (i.e., memory, latency, and energy) of a given HW accelerator. 
Compared to manually designed state-of-the-art CapsNets (i.e., DeepCaps), the NASCaps achieves 79\% latency reduction, 88\% energy reduction, 63\% memory reduction, with only 1\% accuracy reduction.

\subsection{Optimizations for DNN Accelerators} 
\label{Sec_PerformanceAndEnergy_DNNacc}

To efficiently run the generated DNN models on accelerator fabric, optimizations should be applied across the HW architecture, i.e., in the off-chip memory, on-chip memory, and on-chip compute engine.

\begin{figure}[hbtp]
\vspace{-0.3cm}
\centering
\includegraphics[width=\linewidth]{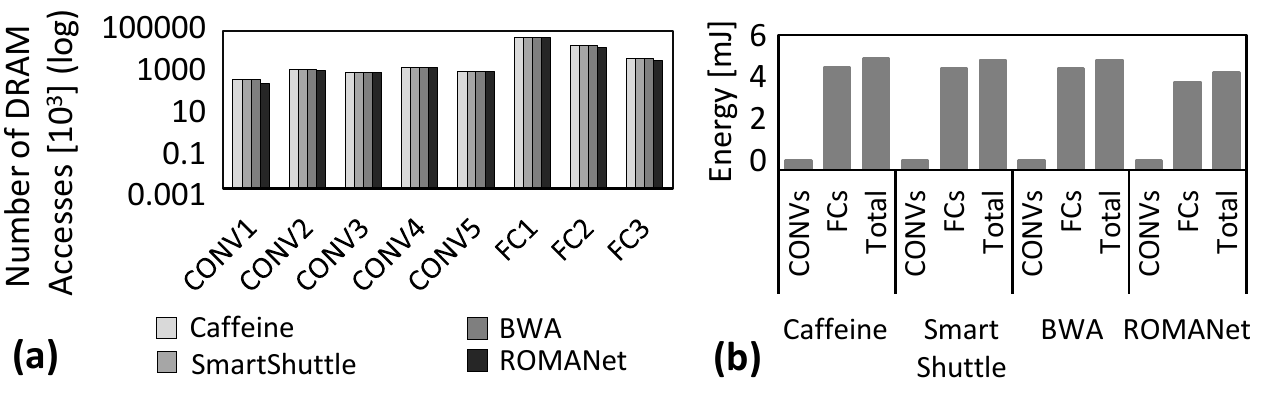}
\vspace{-0.7cm}
\caption{Experimental results of (a) the number of DRAM accesses, and (b) the DRAM access energy for the AlexNet. The ROMANet \cite{Ref_Putra_ROMANet_TVLSI21} decreases the DRAM accesses and the DRAM energy compared to the state-of-the-art (i.e., the Caffeine \cite{Ref_Zhang_Caffeine_TCAD18}, the SmartShuttle \cite{Ref_Li_SmartShuttle_DATE18}, and the BWA \cite{Ref_Tewari_BWA_ISVLSI20}).}
\label{Fig_ROMANet_KeyResults}
\vspace{-0.2cm}
\end{figure}

\begin{figure}[hbtp]
\vspace{-0.3cm}
\centering
\includegraphics[width=\linewidth]{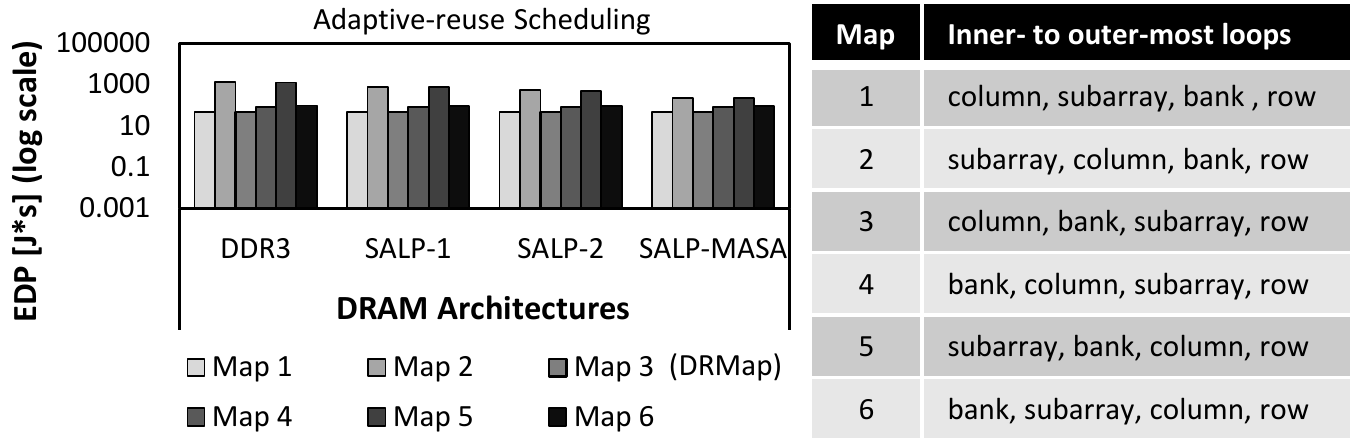}
\vspace{-0.7cm}
\caption{The EDP of DRAM accesses for the AlexNet across different DRAM architectures (i.e., DDR3, SALP-1, SALP-2, and SALP-MASA) and different DRAM mapping policies, which have different orders of DRAM mapping loops. The results show that the DRMap mapping (i.e., Map 3) consistently obtains the lowest EDP \cite{Ref_Putra_DRMap_DAC20}.}
\label{Fig_DRMap_KeyResults}
\vspace{-0.2cm}
\end{figure}

\textbf{The Off-chip Memory (DRAM):}
The main challenges arise from the fact that a full DNN model usually cannot be mapped and processed at once on the accelerator fabric due to limited on-chip memory. Therefore, redundant accesses for the same data to DRAM are required, which restricts the systems from achieving high performance and energy efficiency gains, as DRAM access latency and energy are significantly higher than other operations~\cite{Ref_Sze_DNNsSurvey_ProcIEEE17}.
Toward this, previous works have proposed (1) model compression through pruning  \cite{Ref_Han_DeepCompression_ICLR15,Ref_Li_Pruning_arXiv16,Ref_Marchisio_PruNet_IJCNN18,Ref_Anwar_StructuredPruning_JETC17,Ref_He_AMC_ECCV18} and quantization \cite{Ref_Han_DeepCompression_ICLR15}\cite{Ref_Gupta_LimitedPrecision_ICML15,Ref_Jacob_Quantize_CVPR18,Ref_Marchisio_Q-CapsNets_DAC20}, and (2) data partitioning and scheduling schemes \cite{Ref_Zhang_CNNsFPGA_FPGA15, Ref_Zhang_Caffeine_TCAD18, Ref_Li_SmartShuttle_DATE18, Ref_Tewari_BWA_ISVLSI20}.
However, they do not study the impact of DRAM accesses which dominate the total system latency and energy, 
and do not minimize redundant accesses for overlapping data in convolutional operations.
To address these limitations, several SW-level techniques have been proposed (the ROMANet \cite{Ref_Putra_ROMANet_TVLSI21} and DRMap \cite{Ref_Putra_DRMap_DAC20} methodologies). 
Our ROMANet~\cite{Ref_Putra_ROMANet_TVLSI21} minimizes the DRAM energy consumption through a design space exploration (DSE) that finds the most effective data partitioning and scheduling while considering redundant access optimization. It minimizes the average DRAM energy-per-access by avoiding row buffer conflicts and misses through an effective DRAM mapping, as shown in Fig.~\ref{Fig_ROMANet_KeyResults}.
Our DRMap \cite{Ref_Putra_DRMap_DAC20} further improves the DRAM latency and energy for DNN processing considering different DRAM architectures such as the low-latency DRAM with subarray-level parallelism (i.e., SALP \cite{Ref_Kim_SALP_ISCA12}).
It employs a DSE with a generic DRAM data mapping policy that maximizes DRAM row buffer hits, bank- and subarray-level parallelism to obtain minimum energy-delay-product (EDP) of DRAM accesses for the given DRAM architectures and DNN data partitioning and scheduling (see Fig.~\ref{Fig_DRMap_KeyResults}). 

\begin{figure}[hbtp]
\vspace{-0.2cm}
\centering
\includegraphics[width=\linewidth]{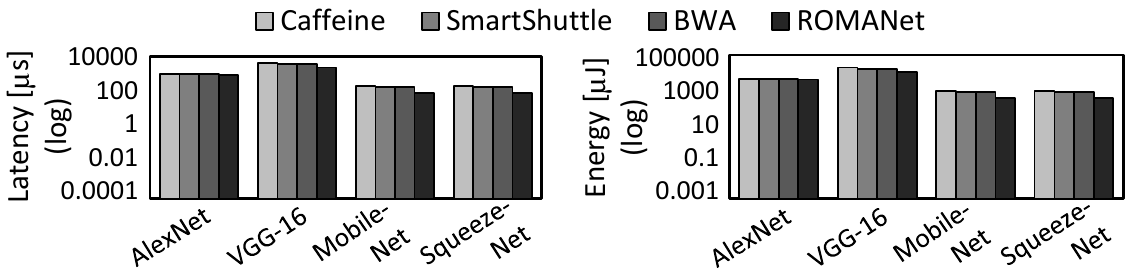}
\vspace{-0.7cm}
\caption{Experimental results of the buffer access latency and energy across different optimization techniques and different networks. The ROMANet \cite{Ref_Putra_ROMANet_TVLSI21} effectively reduces the buffer latency and energy over the state-of-the-art (i.e., the Caffeine \cite{Ref_Zhang_Caffeine_TCAD18}, SmartShuttle \cite{Ref_Li_SmartShuttle_DATE18}, and BWA \cite{Ref_Tewari_BWA_ISVLSI20} techniques).}
\label{Fig_ROMANet_KeyResults_Buff}
\vspace{-0.2cm}
\end{figure}

\smallskip
\textbf{The On-chip Memory (Buffer):}
To efficiently shuttle data between the DRAM and the on-chip fabric, specialized on-chip buffer design and access management are important. 
Here, the scratchpad memory (SPM) design is commonly used due to its low latency and power characteristics~\cite{Ref_Putra_ROMANet_TVLSI21}\cite{Ref_Kang_OnChipMemory_TC21}.
For optimizing buffer access latency and energy, several SW-level techniques have been proposed (such as ROMANet~\cite{Ref_Putra_ROMANet_TVLSI21} and DESCNet~\cite{Ref_Marchisio_DESCNet_TCAD20}). 
Our ROMANet framework~\cite{Ref_Putra_ROMANet_TVLSI21} exploits the bank-level parallelism in the buffer to minimize latency and energy of the given buffer access requests, as shown in Fig.~\ref{Fig_ROMANet_KeyResults_Buff}. 
Meanwhile, our DESCNet framework~\cite{Ref_Marchisio_DESCNet_TCAD20} searches for different on-chip memory architectures to reduce the energy consumption, and performs run-time memory management to power-gate the unnecessary memory blocks for non-memory-intensive operations. These optimizations provide up to 79\% energy savings for CapsNet inference.

\textbf{The Compute Engine (Computational Units):} 
The state-of-the-art HW-level optimization techniques (e.g., approximate computing) can provide significant area, performance and energy efficiency improvements, but at the cost of output quality degradation, which cannot be tolerated in safety-critical applications.
Toward this, we proposed the concept of curable approximations in~\cite{Ref_Hanif_CANN_DAC19}, which ensures minimal accuracy degradation by employing approximations in a way that approximation errors from one stage are compensated in the subsequent stage/s of the pipeline. When used for improving the efficiency of compute engine with cascaded processing elements (PEs), like the systolic array in the TPU \cite{Ref_Jouppi_TPU_ISCA17}, it reduces the Power-Delay Poduct (PDP) of the array by about 46\% and 38\% compared to the conventional and approximate systolic array design, respectively.
To efficiently employ approximations in applications that can tolerate minor quality degradation, a systematic error analysis is necessary to identify the approximation knobs and the degree to which each type of approximation can be employed. 
Toward this, several methodologies have been proposed to analyze the error resilience of CNNs \cite{Ref_Hanif_ErrorResilienceCNN_DATE18} and CapsNets (i.e., ReD-CaNe~\cite{Ref_Marchisio_ReD-CaNe_DATE20}).
By modeling the effects of approximations, it is possible to identify the optimal approximate components (e.g., adders and multipliers) that offer the best accuracy-efficiency trade-off while meeting the user-defined constraints. 
Compared to having accurate hardware, an efficient design that employs a layer-wise selection of approximate multipliers can achieve 28\% energy reduction~\cite{Ref_Marchisio_ReD-CaNe_DATE20}. 
Furthermore, to find the configurations that offer good accuracy-energy trade-offs, the ALWANN framework~\cite{Ref_Mrazek_ALWANN_ICCAD19} performs a DSE with a multi-objective NSGA-II algorithm.

\smallskip
\textbf{Run-time Optimizations:} 
Several run-time power management techniques can be employed to further boost the efficiency, e.g., the run-time clock gating, power gating, and dynamic voltage and frequency scaling (DVFS) techniques. 
For instance, the DESCNet technique~\cite{Ref_Marchisio_DESCNet_TCAD20} partitions the SPM into multiple sectors, and performs sector-level power-gating based on the characteristics of CapsNet workload to get high energy savings at run time during inference. 
Compared to the standard memory designs, the application-driven memory organizations equipped with memory power management unit in the DESCNet can save up to 79\% energy and 47\% area.

\section{Improving Reliability and Security for Edge AI}
\label{Sec_ReliabilityAndSecurity}

\begin{figure*}[t] 
\centering
\includegraphics[width=\linewidth]{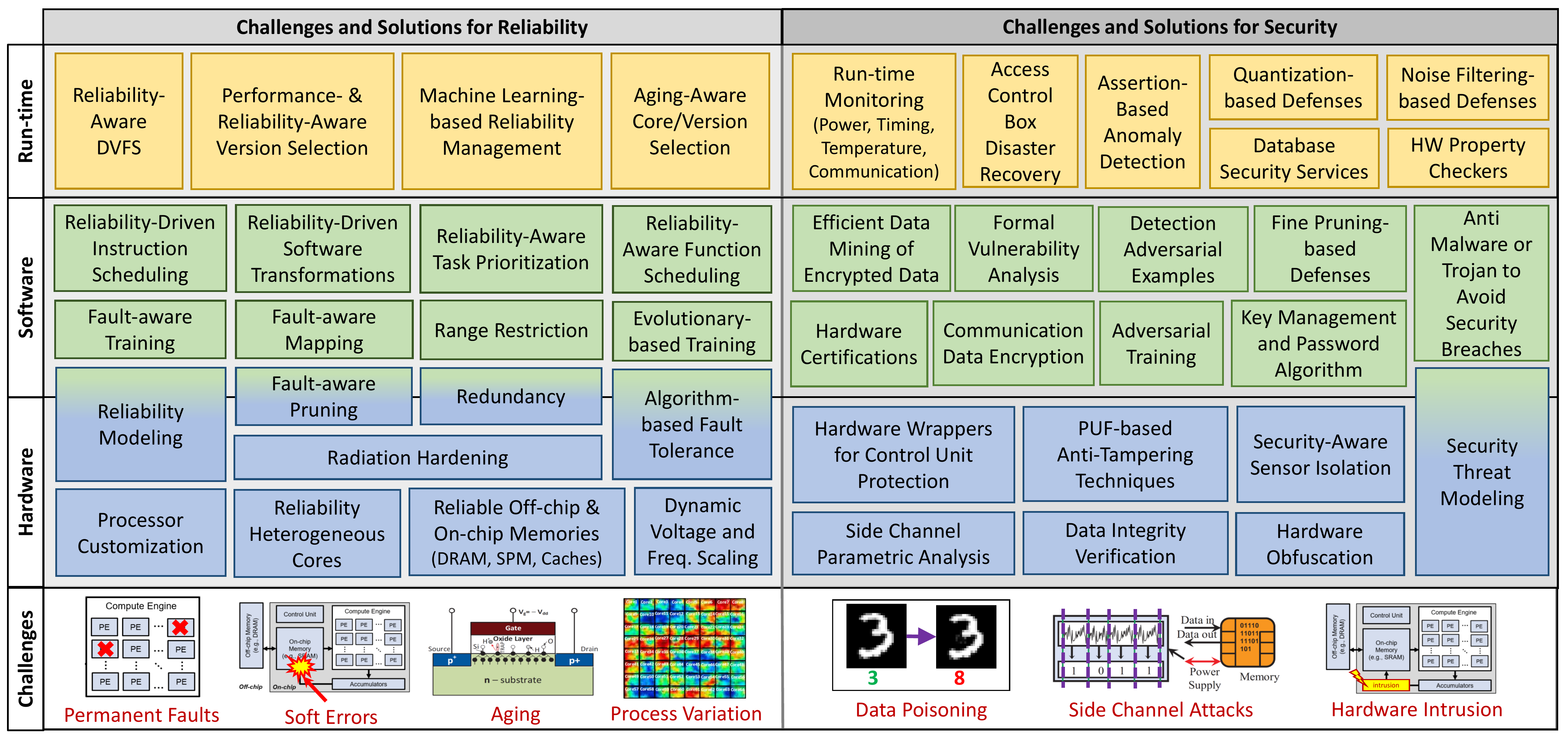}
\vspace{-0.6cm}
\caption{Overview of challenges for reliability and security aspects, and the respective solutions on different system layers.}
\label{Fig_Challenges_ReliableSecurity}
\vspace{-0.4cm}
\end{figure*}

The Edge AI systems need to continuously produce correct outputs under diverse operating conditions.
This requirement is especially important for safety-critical applications such as medical data analysis and autonomous driving. 
There are mainly two categories of vulnerability issues that threaten the Edge AI: (1)~\textit{reliability} and (2)~\textit{security}.
In this section, we discuss the state-of-the-art techniques for improving the reliability and security of Edge AI (see an overview in Fig.~\ref{Fig_Challenges_ReliableSecurity}).

\subsection{Reliability Threats and Mitigation Techniques}

Reliability threats may come from various sources like \textit{process variation}, \textit{soft errors}, and \textit{aging}. 
They can manifest as \textit{permanent  faults} (faults that remain in the system permanently and do not disappear), \textit{transient faults} (faults that occur once and can disappear), or \textit{performance degradation} (e.g., in the form of delay/timing errors).
To address these threats, conventional fault-mitigation techniques for VLSI can be employed, e.g., Dual Modular Redundancy (DMR) \cite{Ref_Vadlamani_DMR_DATE20}, Triple Modular Redundancy (TMR) \cite{Ref_Lyons_TMR_IBM62}, and Error Correction Code (ECC) \cite{Ref_Sze_ECCs_USPatent00}. 
However, these techniques incur huge overheads due to redundant hardware or execution.
Hence, \textit{cost-effective techniques are required to mitigate the reliability threats in the Edge AI}.

\smallskip
\textbf{Permanent Faults:} 
To mitigate permanent faults in DNN accelerators, recent works have proposed techniques like fault-aware pruning (FAP) \cite{Ref_Zhang_FAP_VTS18} and fault-aware training (FAT) \cite{Ref_Zhang_FAP_VTS18}\cite{Ref_Kim_FAT_TCASI18}.
They aim at making DNNs resilient to the faults by incorporating the information of faults in the optimization/training process. 
These techniques usually require minor modifications at the hardware level (i.e., additional circuitry) to bypass/disconnect the faulty components, which results in minor run-time overheads.   
The key limitation of FAT is that it incurs a huge retraining cost, specifically for the cases in which retraining has to be performed for a large number of faulty chips. Moreover, FAT cannot be employed if the training dataset is not available to the user.
To address these limitations, we proposed SalvageDNN~\cite{Ref_Hanif_SalvageDNN_RSTA20} that enables us to mitigate permanent faults in DNN accelerators without retraining. 
It achieves this through a significance-driven fault-aware mapping (FAM) strategy, and shuffling of parameters at the software level to avoid additional memory operations. 
%
Techniques like FT-ClipAct~\cite{Ref_Hoang_FT-ClipAct_DATE20} and Ranger~\cite{Ref_Chen_Ranger_DSN21} employ range restriction functions to block large (abnormal) activation values using pre-computed thresholds. 
Range restriction is realized using clipped activation functions that map out of the range values to pre-specified values within the range that have the least impact on the output. 
FT-ClipAct~\cite{Ref_Hoang_FT-ClipAct_DATE20} shows that such techniques can improve the accuracy of the VGG-16 by 68.92\% (on average) at 10$^{-5}$ fault rate compared to the no fault mitigation case. 

\smallskip
\textbf{Transient Faults (Soft Errors):} 
Soft error rates have been increasing in HW systems~\cite{Ref_Li_SoftErrorDNN_SC17}.
To mitigate their negative impact, several techniques have been proposed~\cite{Ref_Schorn_OnlineErrorDet_Safecomp18,Ref_Ozen_SanityCheck_ATS18,Ref_Hong_TerminalBrainDamage_USENIX19,Ref_Mahmoud_HarDNN_arXiv20,Ref_Zhao_FTCNN_TPDS21,Ref_Chen_Ranger_DSN21}. 
Some of these techniques only cover limited faults~\cite{Ref_Hong_TerminalBrainDamage_USENIX19} and/or incur significant overheads~\cite{Ref_Schorn_OnlineErrorDet_Safecomp18}\cite{Ref_Mahmoud_HarDNN_arXiv20}. 
For instance, techniques in~\cite{Ref_Schorn_OnlineErrorDet_Safecomp18} employ a separate network to detect the anomaly in the output. 
Other state-of-the-art techniques employ online SW-level range restriction functions, like Ranger~\cite{Ref_Chen_Ranger_DSN21} that rectifies the faulty outputs of DNN operations without re-computation by restricting the value ranges.

\smallskip
\textbf{Aging:}
Aging may result in timing errors, and techniques like ThUnderVolt~\cite{Ref_Zhang_Thundervolt_DAC18} and GreenTPU~\cite{Ref_Pandey_GreenTPU_DAC19} can be employed for mitigating the effects of timing errors that occur in the computational units of DNN accelerators.
Meanwhile, aging in the on-chip memory (6T-SRAM), one of the key component in DNN accelerators, has been addressed by techniques like the fixed aging balancing~\cite{Ref_Kumar_NBTIonSRAM_ISQED06}, adaptive aging balancing~\cite{Ref_Shafique_EnAAM_DAC15}, and additional circuitry~\cite{Ref_Siddiqua_NBTIinSRAM_TVLSI11}\cite{Ref_Zatt_LoPoMemArch_ICCAD11}.
However, these techniques are designed for specific data distribution and/or applications, or require additional circuitry in each SRAM cell. 
To address this challenge, we proposed DNN-Life framework~\cite{Ref_Hanif_DNNlife_DATE21} that employs novel memory-write (and read) transducers to achieve an optimal duty-cycle at run time in each cell of the on-chip weight memory to mitigate NBTI aging. 

\smallskip
Besides the HW-induced reliability threats (i.e., permanent faults, soft errors, and aging), other works have analyzed the resilience of DNNs against other threats (e.g., input noise). 
For instance, the FANNet methodology~\cite{Ref_Naseer_FANNet_DATE20} analyzes the DNN noise tolerance using model checking techniques for formal analysis of DNNs under different ranges of input noise.  
The key idea is to investigate the impact of training bias on accuracy, and study the input node sensitivity under noise.

\subsection{Secure ML: Attacks and Defenses}

\begin{figure}[hbtp]
\centering
\includegraphics[width=\linewidth]{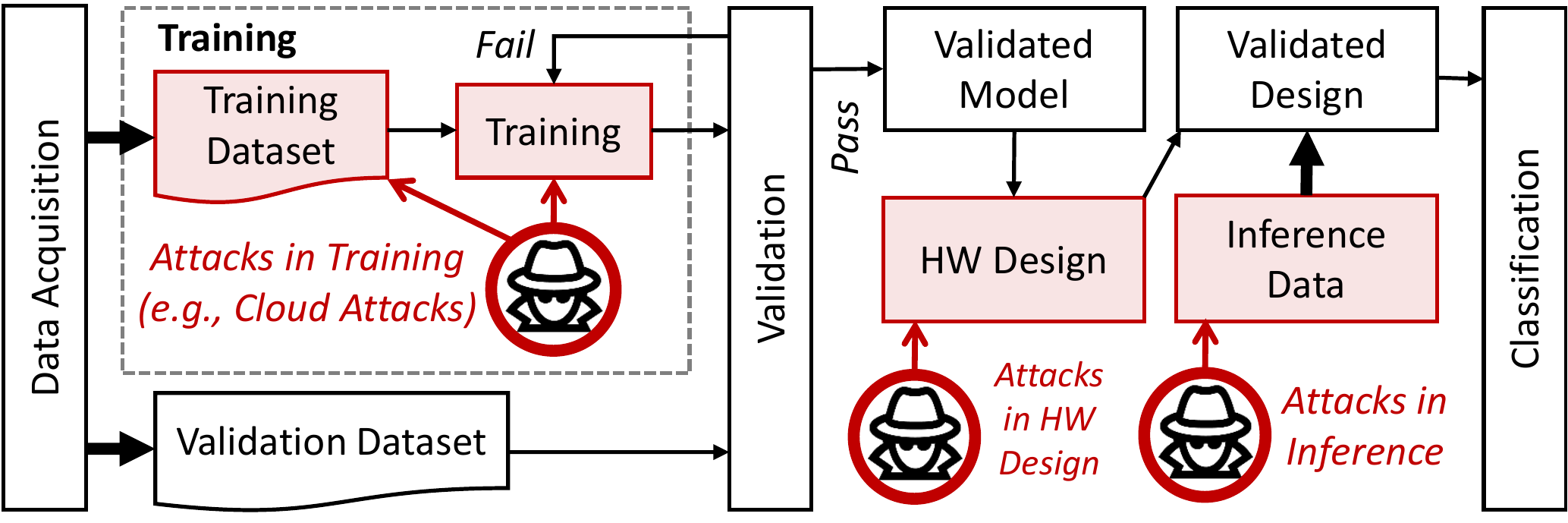}
\vspace{-0.6cm}
\caption{An overview of security threats (attacks) and in the training, HW design/implementation, and inference phases.}
\label{Fig_Overview_SecurityThreats}
\vspace{-0.6cm}
\end{figure}

Security threats may come from different types of attacks, such as \textit{side-channel attacks}, \textit{data poisoning}, and \textit{hardware intrusion}. 
These attacks can cause confidence reduction in classification accuracy, random or targeted misclassification, and IP stealing.
To systematically identify the possible security attacks and defense mechanisms for Edge AI, \textit{a threat model} (which defines the capabilities and goals of the attacker under realistic assumptions) is required \cite{Ref_Hanif_RobustML_IOLTS18}. 
The attacks can be categorized based on the Edge AI design cycle, i.e., during \textit{training}, \textit{HW design or implementation}, and \textit{inference} \cite{Ref_Hanif_RobustML_IOLTS18} (the overview is shown in Fig.~\ref{Fig_Overview_SecurityThreats}).

\begin{itemize}[leftmargin=*]
    \item \textbf{Training:}
    The attacker can manipulate the DNN model, training dataset or tools, to attack the system \cite{Ref_Papernot_SecurityPrivacyML_arXiv16}.
    \item \textbf{HW Implementation:}
    The attacker can steal the DNN IP through side-channel attacks, or hardware intrusion~\cite{Ref_Papernot_SecurityPrivacyML_arXiv16}.
    \item \textbf{Inference:}
    The attacker can perform side-channel attacks for IP stealing, or manipulate the input data to achieve random or targeted misclassification~\cite{Ref_Papernot_SecurityPrivacyML_arXiv16}.
\end{itemize}
Therefore, \textit{effective defense mechanisms are required to secure Edge AI from possible attacks}.
Toward this, both attacks and defenses need to be explored.
In this section, we discuss different security attacks and some possible defenses (countermeasures) against these attacks. 

\smallskip
\textbf{Data Poisoning/Manipulation:} 
Data poisoning aims at producing incorrect output (i.e., misclassification), and it can be performed by adding  crafted noise to the DNN inputs (i.e., training or test data).
Toward this, SW-level methodologies (e.g., TrISec~\cite{Ref_Khalid_TrISeC_IOLTS19}, FaDec~\cite{Ref_Khalid_FaDec_IJCNN20}, and CapsAttacks~\cite{Ref_Marchisio_CapsAttacks_arxiv19}) have been proposed to explore the impacts of different data poisoning attacks.
For instance, TrISec~\cite{Ref_Khalid_TrISeC_IOLTS19} generates imperceptible attack images as the test inputs by leveraging the backpropagation algorithm on the trained DNNs without knowledge of the training dataset. 
The generated attacks have close correlation and structural similarity index with the clean input, thereby making them difficult to notice in both subjective and objective tests.
FaDec~\cite{Ref_Khalid_FaDec_IJCNN20} generates imperceptible decision-based attack images as the test inputs by employing fast estimation of the classification boundary and adversarial noise optimization. 
It results in a fast and imperceptible attack, i.e., 16x faster than the state-of-the-art decision-based attacks.
Meanwhile, CapsAttacks~\cite{Ref_Marchisio_CapsAttacks_arxiv19} performs analysis to study the vulnerabilities of the CapsNet by adding perturbations to the test inputs. The results show that, compared to traditional DNNs with similar width and depth, the CapsNets are more robust to affine transformations and adversarial attacks.
\textit{All these works demonstrated that DNNs are vulnerable to data poisoning attacks (which can be imperceptible), thereby the effective countermeasures are required}. 
Previous works have proposed several SW-level defense mechanisms. 
One idea is to employ encryption for protecting the training data \cite{Ref_Hesamifard_CryptoDL_arXiv17,Ref_Gonzalez_EncryptedTrainData_IJIS18,Ref_Gao_PANDA_arXiv18,Ref_Hesamifard_Privacy_PET18}. 
Another idea is to employ noise filters, as the FadeML methodology~\cite{Ref_Khalid_FadeML_DATE19} demonstrates that the existing adversarial attacks can be nullified using noise filters, like the Local Average with Neighborhood Pixels (LAP) and Local Average with Radius (LAR) techniques.
Meanwhile, the QuSecNets methodology~\cite{Ref_Khalid_QuSecNets_IOLTS19} employs quantization to eliminate the attacks in the input images. 
It has two quantization mechanisms, i.e., \textit{constant quantization}, which quantizes the intensities of input pixels based on fixed quantization levels; and \textit{trainable quantization}, which learns the quantization levels during the training phase to provide a stronger protection.
This technique increases the accuracy of CNNs by 50\%-96\% and by 10\%-50\% for the perturbed images from the MNIST and the CIFAR-10, respectively.

\smallskip
\textbf{Side Channel Attacks:} 
These attacks aim at extracting confidential information (e.g., for data sniffing and IP stealing) without interfering with the functionality or the operation of the devices  by monitoring and manipulating the side channel parameters (e.g., timing, power, temperature, etc.).
The potential countermeasures are the obfuscation techniques, which target at concealing or obscuring the functional behavior or specific information \cite{Ref_Hoque_HWobfuscation_DnT20}. 
For instance, the processing HW can be designed so that the power signals of the operation are independent to the processed data values, thereby concealing the secret information \cite{Ref_Bhunia_SCA_HWSecurity19}. 
Meanwhile, to protect the devices from timing attacks, designers can (1) randomize the execution delay of different operations, or (2) enforce the same execution delay for all operations, thereby obscuring the underlying operation \cite{Ref_Bhunia_SCA_HWSecurity19}.

\smallskip
\textbf{Hardware Intrusion:} 
HW intrusion means that the attacker inserts malware or trojan (typically in the form of circuitry modification) in the processing HW for performing attacks such as confidence reduction and misclassification.
The potential countermeasures are the typical HW security techniques, like the built-in self-test (BIST) to verify the functionality of the processing HW, the side channel analysis-based monitoring  \cite{Ref_Huang_SafetyDNNs_CAV17,Ref_Hunt_Chiron_arXiv18,Ref_Wei_IKnowWhatYouSee_ACSAC18} to detect and identify anomalous side channel signals, the formal method analysis to quickly and comprehensively analyze the behavior of the processing HW (e.g., using property checker \cite{Ref_Huang_SafetyDNNs_CAV17}, mathematical model \cite{ref_Gopinath_Deepsafe_ATVA18}, SAT solver \cite{Ref_Kuper_Verification_arXiv18}, and SMT solver \cite{Ref_Katz_Reluplex_CAV17}). 

\section{A Cross-Layer Framework for \\ Energy-Efficient and Robust Edge AI}
\label{Sec_IntegratedFrameworks}

\begin{figure*}[t]
\centering
\includegraphics[width=\linewidth]{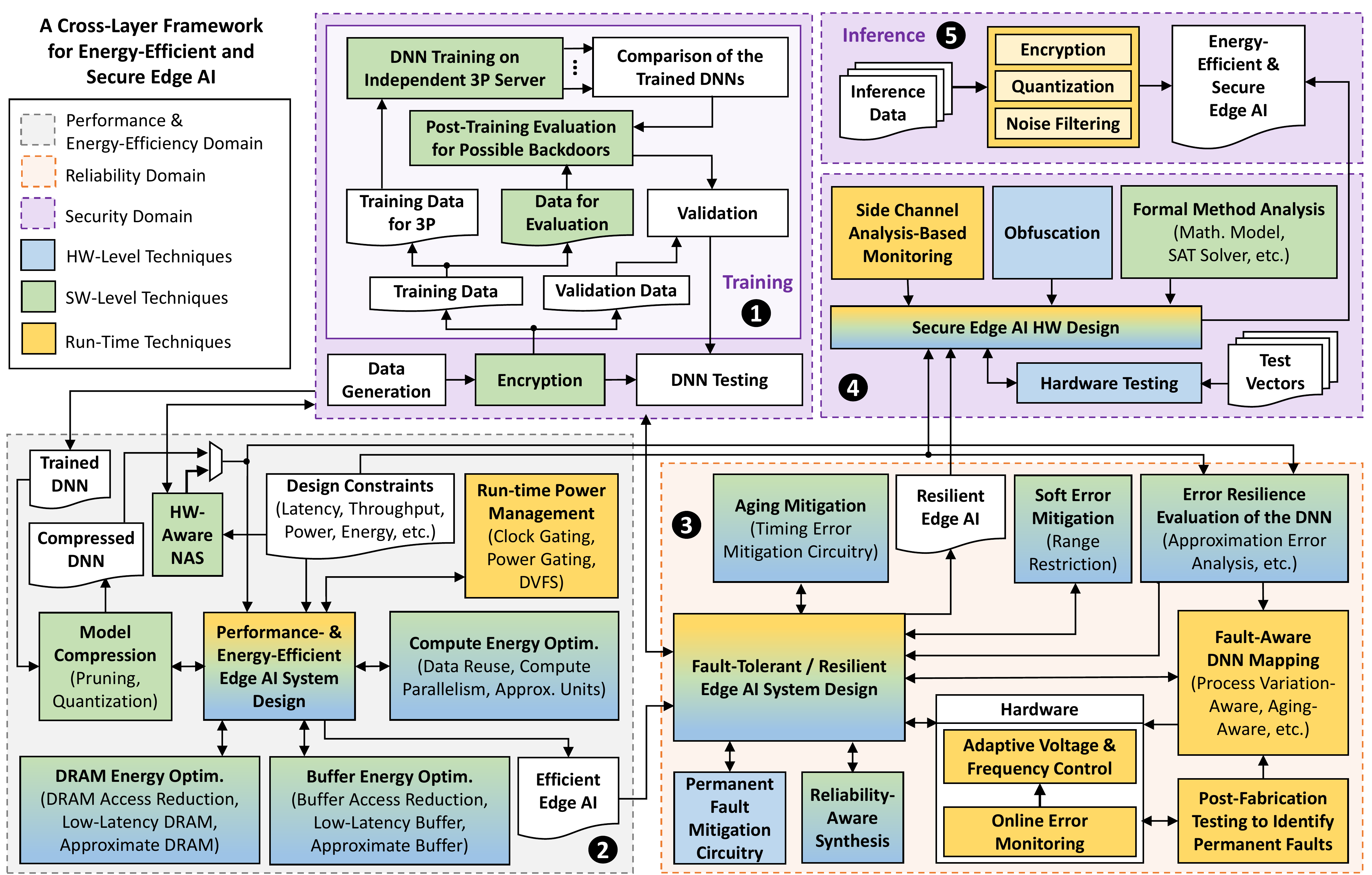}
\vspace{-0.6cm}
\caption{Overview of our cross-layer framework for energy-efficient and secure Edge AI systems.}
\label{Fig_IntegratedFrameworks}
\vspace{-0.4cm}
\end{figure*}

To develop energy-efficient and robust Edge AI systems, different aspects related to performance and energy efficiency, reliability, and security should be collectively addressed. 
Toward this, we propose a cross-layer framework that combines different techniques from different layers of the computing stack for achieving energy-efficient and secure Edge AI systems (see the overview in Fig.~\ref{Fig_IntegratedFrameworks}). 
Our integrated framework employs the following steps.

\smallskip
\textbf{DNN Model Creation with Secure Training:} 
DNNs for Edge AI have to meet the design constraints (e.g., accuracy, memory, power, and energy). This can be achieved through two different ways, i.e., by employing (1) model compression through pruning \cite{Ref_Marchisio_PruNet_IJCNN18} and quantization \cite{Ref_Marchisio_Q-CapsNets_DAC20} of the pre-trained DNN model, and (2) multi-objective neural architecture search (NAS) similar to the APNAS~\cite{Ref_Achararit_APNAS_Access20} and NASCaps~\cite{Ref_Marchisio_NASCaps_ICCAD20} frameworks. 
APNAS \cite{Ref_Achararit_APNAS_Access20} searches for a model that has good accuracy and performance considering a systolic array-based DNN accelerator \cite{Ref_Hanif_MPNA_arXiv18} through reinforcement learning.
Meanwhile, NASCaps~\cite{Ref_Marchisio_NASCaps_ICCAD20} optimizes the accuracy and the hardware efficiency of a given accelerator for CapsNet inference. 
%
%
To ensure that the generated model can be trusted, the training process should be protected from attacks. 
To do this, several countermeasures can be employed, e.g., by comparing the redundant trained models~\cite{Ref_Li_OutsourcedService_JNCA18}, by performing local training~\cite{Ref_Liu_LessIsMore_GameSec18} to identify if the trained model has been attacked, or by encrypting the training dataset~\cite{Ref_Gao_PANDA_arXiv18, Ref_Hesamifard_Privacy_PET18,Ref_Hesamifard_CryptoDL_arXiv17,Ref_Gonzalez_EncryptedTrainData_IJIS18} to remove data poisoning attacks (see \circledB{1} in Fig. \ref{Fig_IntegratedFrameworks}).

\smallskip
\textbf{Efficient Edge AI Design:} 
Once a trusted model is generated, further performance and energy optimizations are performed (see \circledB{2} in Fig. \ref{Fig_IntegratedFrameworks}). 
At design time, DRAM latency and energy can be improved using techniques like ROMANet  \cite{Ref_Putra_ROMANet_TVLSI21} and DRMap \cite{Ref_Putra_DRMap_DAC20}. 
Meanwhile, the buffer latency and energy can be optimized using ROMANet \cite{Ref_Putra_ROMANet_TVLSI21} and DESCNet \cite{Ref_Marchisio_DESCNet_TCAD20}, and the compute latency and energy can be optimized using approximation methodologies like CANN \cite{Ref_Hanif_CANN_DAC19}, ALWANN \cite{Ref_Mrazek_ALWANN_ICCAD19}, and ReD-CaNe \cite{Ref_Marchisio_ReD-CaNe_DATE20}.
Moreover, efficiency gains of the systems can be improved at run time using run-time power management techniques like clock gating \cite{Ref_Kriebel_RobustnessML_ISVLSI18}, power gating \cite{Ref_Marchisio_DESCNet_TCAD20}, and DVFS\cite{Ref_Kriebel_RobustnessML_ISVLSI18}.
Furthermore, this step should ensure that the employed techniques do not violate the design specifications, thereby providing efficient Edge AI. 

\smallskip
\textbf{Resilient Edge AI Design:} 
To improve the resiliency of Edge AI against the reliability threats, effective mitigation techniques are required (see \circledB{3} in Fig. \ref{Fig_IntegratedFrameworks}). 
Toward this, the characteristics of DNN resiliency under the targeted reliability threats are evaluated. 
Recent works have studied the DNN resiliency in the presence of approximation errors \cite{Ref_Hanif_ErrorResilienceCNN_DATE18}\cite{Ref_Marchisio_ReD-CaNe_DATE20} and permanent faults~\cite{Ref_Hanif_SalvageDNN_RSTA20}.
Based on this information, appropriate fault mitigation techniques can be identified and deployed. 
At design time, several techniques can be employed, such as fault-aware training (e.g., FAP \cite{Ref_Zhang_FAP_VTS18} and FAT \cite{Ref_Kim_FAT_TCASI18}), range restriction (e.g., FT-ClipAct \cite{Ref_Hoang_FT-ClipAct_DATE20}), and aging-aware timing error mitigation (e.g., ThUnderVolt \cite{Ref_Zhang_Thundervolt_DAC18} and GreenTPU \cite{Ref_Pandey_GreenTPU_DAC19}).  
Meanwhile, the fault-aware mapping (e.g., SalvageDNN \cite{Ref_Hanif_SalvageDNN_RSTA20}), the range restriction (e.g., Ranger \cite{Ref_Chen_Ranger_DSN21}), online error monitoring and adaptive DVFS can be performed to improve the system's resiliency at run time.
Furthermore, this step needs to ensure that the employed techniques do not lead to any violation of the design constraints, thereby resulting in a resilient and energy-efficient Edge AI system.

\smallskip
\textbf{Secure HW Design/Implementation:} 
Since the HW side also has vulnerability issues, the HW design/implementation process should be protected. 
Toward this, the existing HW security techniques can be employed (see \circledB{4} in Fig. \ref{Fig_IntegratedFrameworks}). 
For instance, the side-channel analysis-based monitoring~\cite{Ref_Huang_SafetyDNNs_CAV17,Ref_Hunt_Chiron_arXiv18,Ref_Wei_IKnowWhatYouSee_ACSAC18} can monitor the side-channel signals that attackers can exploit. 
Then, we can leverage the information to devise defense mechanisms that block the exploitation. 
Another idea is to obscure the HW information from the attacker using obfuscation techniques \cite{Ref_Rosenberg_DeepAPT_ICANN17}\cite{Ref_Sun_Obfuscation_CVPR18}. 
The other techniques leverage the formal method-based analysis \cite{Ref_Huang_SafetyDNNs_CAV17}\cite{ref_Gopinath_Deepsafe_ATVA18,Ref_Kuper_Verification_arXiv18,Ref_Katz_Reluplex_CAV17} to quickly identify all possible security vulnerabilities and the corresponding defense mechanisms. 
To evaluate the efficacy of the applied defense mechanisms, HW testing is performed. 
Furthermore, this step also needs to ensure that the employed defense techniques still meet the design constraints, thereby resulting in a secure HW design.

\smallskip
\textbf{Secure Inference:} 
Since the security attacks can also target the inference phase, a secure inference is required (see \circledB{5} in Fig. \ref{Fig_IntegratedFrameworks}).
Most of the attacks come in the form of data manipulation. 
Hence, we can perform data encryption to block the insertion of perturbations into the input data. 
Another idea is to mitigate the input data-based attacks by employing quantization-based defenses such as, QuSecNets~\cite{Ref_Khalid_QuSecNets_IOLTS19} and by noise filters like in the FadeML methodology~\cite{Ref_Khalid_FadeML_DATE19}. 

\smallskip
Note that all the proposed steps can jointly provide an end-to-end cross-layer framework that performs HW- and SW-level optimizations at the design-time and run-time. 
\textit{Our proposed framework ensures that the Edge AI systems have high performance and energy efficiency, while providing correct output under diverse reliability and security threats}.

\section{Neuromorphic Research considering SNNs}
\label{Sec_NeuromorphicSNNs}

SNNs are considered as the third generation of NN models, which employ spike-encoded information and computation \cite{Ref_Maass_SNNs_NN97}. 
Due to their bio-inspired operations, SNNs have a high potential to provide energy-efficient computation. 
Recent works have been actively exploring two research directions, i.e., SNNs with a localized learning rule like the spike-timing-dependent plasticity (STDP) \cite{Ref_Putra_FSpiNN_TCAD20}, and SNNs obtained from DNN conversions~\cite{Ref_Massa_EfficientSNN_IJCNN20}. 

\subsection{Improving the Energy Efficiency of SNNs}

\begin{figure}[t]
\centering
\includegraphics[width=\linewidth]{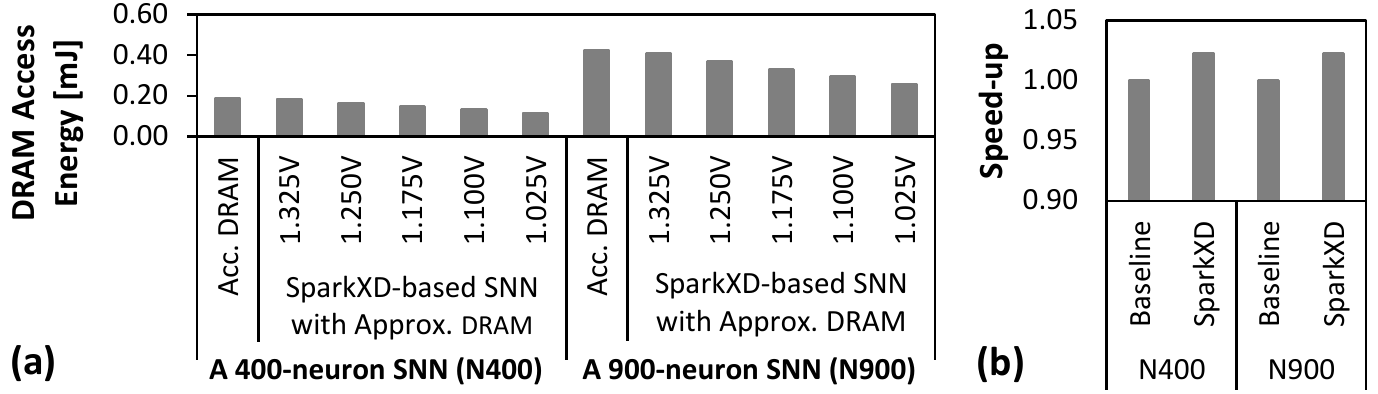}
\vspace{-0.6cm}
\caption{(a) The DRAM energy in an SNN inference on MNIST incurred by an SNN with accurate DRAM (the Baseline) and a SparkXD-based SNN with approximate
DRAM. (b) The speed-up achieved by the SparkXD.}
\label{Fig_SparkXD_KeyResults}
\end{figure}

\begin{figure}[t]
\centering
\includegraphics[width=\linewidth]{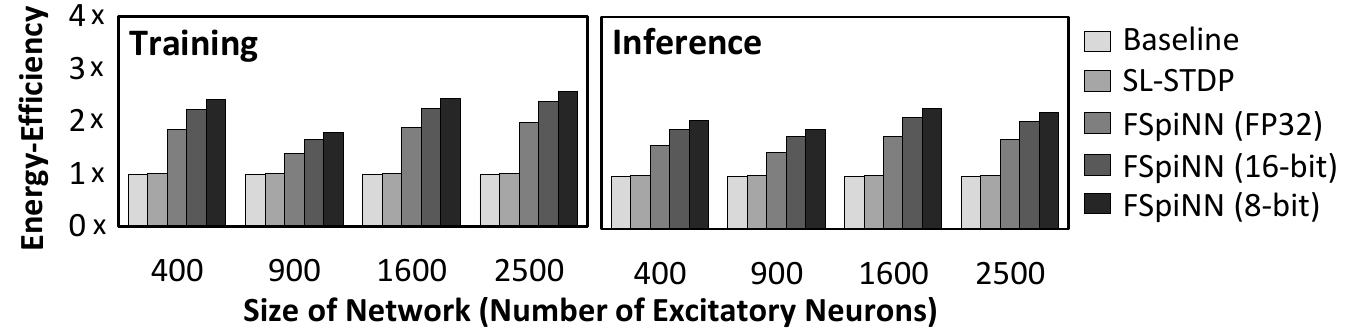}
\vspace{-0.6cm}
\caption{The FSpiNN improves the energy efficiency compared to the standard unsupervised SNN (Baseline) \cite{Ref_Diehl_UnsupervisedSNN_FNCOM15} and the SL-STDP \cite{Ref_Srinivasan_SLSTDP_IJCNN17} across different network sizes for both training and inference phases on the MNIST workload.}
\label{Fig_FSpiNN_KeyResults}
\vspace{-0.4cm}
\end{figure}

\begin{figure}[t]
\centering
\includegraphics[width=0.92\linewidth]{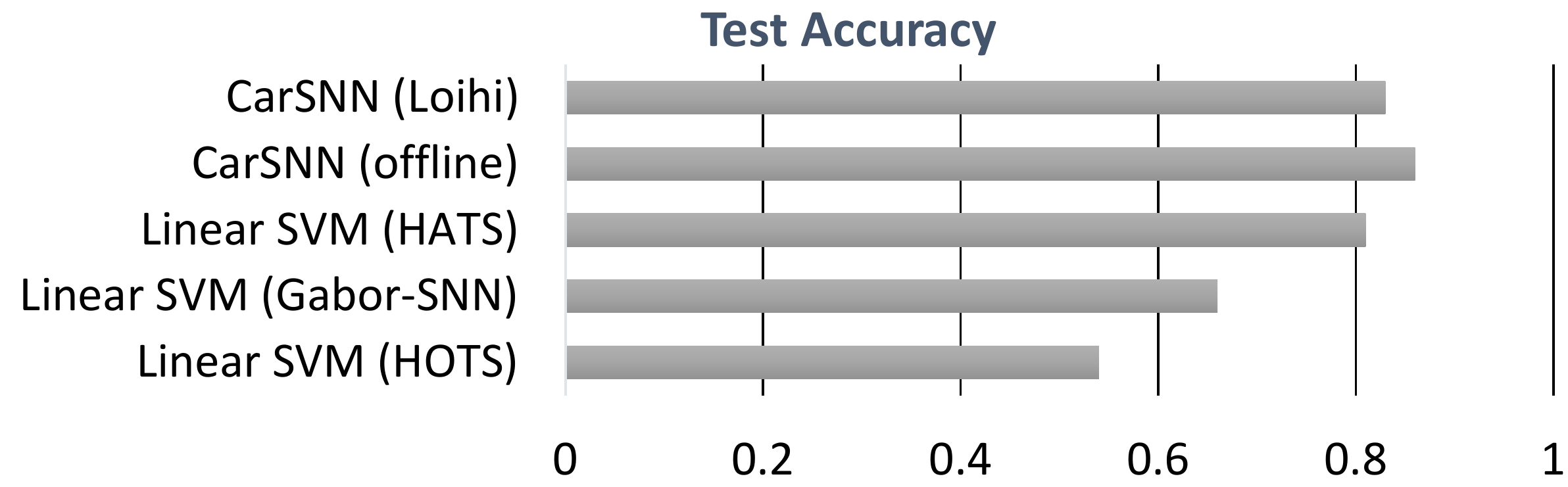}
\vspace{-0.2cm}
\caption{The CarSNN~\cite{Ref_Viale_CarSNN_IJCNN21}, although being more energy-efficient, achieves higher accuracy for the N-CARS dataset than the related works like the HATS~\cite{Ref_Sironi_HATS_CVPR18}, Gabor-SNN~\cite{Ref_Bovik_GaborSNN_TPAMI90}, and HOTS~\cite{Ref_Lagorce_HOTS_TPAMI17} techniques.}
\label{Fig_CarSNN_KeyResults}
\vspace{-0.2cm}
\end{figure}

To improve the energy efficiency of SNNs, several HW- and SW-level optimizations have been proposed.
For HW-level techniques, SNN accelerators have been designed, such as TrueNorth~\cite{Ref_Akopyan_TrueNorth_TCAD15}, SpiNNaker~\cite{Ref_Painkras_SpiNNaker_JSCC13}, PEASE~\cite{Ref_Roy_PEASE_ISLPED17}, Loihi~\cite{Ref_Davies_Loihi_IEEEMICRO18}, and ODIN~\cite{Ref_Frenkel_ODIN_TBCAS18}. 
%
Recent work (i.e., the SparkXD framework~\cite{Ref_Putra_SparkXD_DAC21}) optimizes the DRAM access latency and energy for SNN inference by employing the reduced-voltage DRAM operations and effective DRAM mapping, leading to DRAM energy saving by up to 40\% (see Fig.~\ref{Fig_SparkXD_KeyResults}).  
For SW-level techniques, the FSpiNN framework~\cite{Ref_Putra_FSpiNN_TCAD20} improves the energy efficiency of SNN processing in the training (avg. 3.5x) and the inference (avg. 1.8x) through the optimization of neural operations and quantization, without accuracy loss (see Fig.~\ref{Fig_FSpiNN_KeyResults}). 
The Q-SpiNN~\cite{Ref_Putra_QSpiNN_IJCNN21} explores different precision levels, rounding schemes, and quantization schemes (i.e., post- and in-training quantization) to maximize memory savings for both weights and neuron parameters (which occupy considerable amount of memory in the accelerator fabric).
The other techniques target at mapping and running the SNN applications (e.g., DVS Gesture Recognition \cite{Ref_Massa_EfficientSNN_IJCNN20} and Autonomous Cars \cite{Ref_Viale_CarSNN_IJCNN21}) on neuromorphic hardware (i.e., Loihi) to improve the energy efficiency of their processing compared to running them on conventional platforms (e.g., CPUs, GPUs). 
As shown in Fig.~\ref{Fig_CarSNN_KeyResults}, the CarSNN~\cite{Ref_Viale_CarSNN_IJCNN21} improves by 2\% the N-CARS accuracy, compared to the related works, while consuming only 315 mW on the Loihi Neuromorphic Chip, thus making a step forward towards ultra-low power event-based vision for autonomous cars.

\subsection{Improving the Reliability of SNNs}

\begin{figure}[t]
\centering
\includegraphics[width=\linewidth]{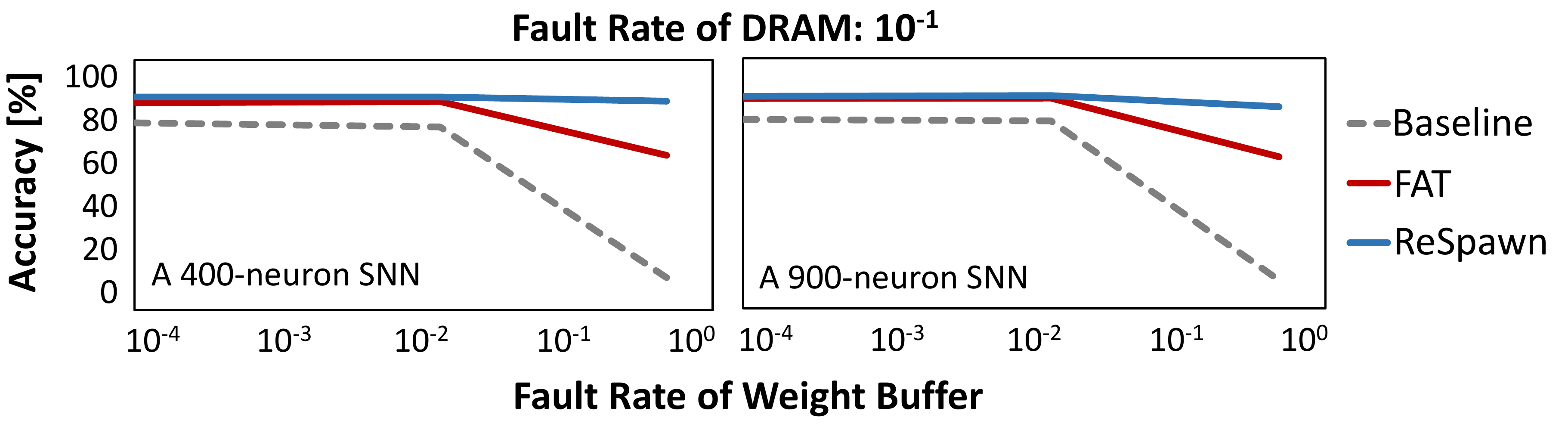}
\vspace{-0.6cm}
\caption{The ReSpawn maintains higher accuracy than the fault-aware training (FAT), across different network sizes and different fault rates in memories.}
\label{Fig_ReSpawn_KeyResults_Acc}
\vspace{-0.4cm}
\end{figure}

In recent years, the SNN reliability aspect starts gaining attention because it is crucial to ensure the functionality of SNN systems. 
Moreover, the reliability issues may come from various sources (e.g., manufacturing defects, optimization techniques, etc.).  
For instance, employing the reduced-voltage DRAM in SNN accelerators can offer energy savings, but at the cost of increased DRAM errors which may alter the weight values and reduce the accuracy. 
Toward this, the SparkXD framework~\cite{Ref_Putra_SparkXD_DAC21} improves the SNN reliability (preserving the high accuracy) by incorporating the information of faults (i.e., fault map and fault rate) in the retraining process, i.e., so-called the fault-aware training (FAT).  
Furthermore, the ReSpawn framework~\cite{Ref_Putra_ReSpawn_ICCAD21} mitigates the negative impact of permanent and approximation-induced faults in the off-chip and on-chip memories of SNN HW accelerators through a cost-effective fault-aware mapping (FAM).  
It places the weight bits with higher significance on the non-faulty memory cells, which enhances the reliability of SNNs without retraining, and achieves up to 70\% accuracy improvement from the baseline, as shown in Fig.~\ref{Fig_ReSpawn_KeyResults_Acc}.
In this manner, the ReSpawn can also improve the yield and reduce the per-unit-cost of SNN chips. 
Besides the HW-induced faults, the SNN systems may encounter dynamically changing environments, which cause the offline-learned knowledge to obsolete at run-time. 
Toward this, the SpikeDyn framework~\cite{Ref_Putra_SpikeDyn_DAC21} employs an unsupervised continual learning mechanism by leveraging the internal characteristics of neural dynamics and weight decay function to enable an online learning scenario. 


\subsection{Improving the Security of SNNs}

Previous works have studied that SNNs are vulnerable to security attacks, like data poisoning attacks on traditional image classification datasets like the MNIST~\cite{Ref_Marchisio_IsSpikingSecure_IJCNN20} and on event-based datasets~\cite{Ref_Marchisio_DVS-Attacks_IJCNN21}, showing different behavior under attack, compared to the non-spiking DNNs. 
Furthermore, SNNs are also vulnerable to externally triggered bit-flip attacks. The experiments conducted in~\cite{Ref_Venceslai_NeuroAttack_IJCNN20} show that only 4 bit-flips at the most sensitive weight memory cells are sufficient for fooling SNNs on the CIFAR10 dataset. Once these memory locations are found, the attacker can trigger the malicious hardware that generates bit-flips by inserting a specific pattern in the input images.
To address the security problem, several defense techniques have been proposed. 
One technique is exploiting the structural network parameters, e.g., threshold voltage and time window, to improve the SNN robustness~\cite{Ref_ElAllami_SecuringSNN_DATE21}. By fine-tuning such parameters, the SNNs can be up to 85\% more robust than non-spiking DNNs.
Meanwhile, the R-SNN methodology~\cite{Ref_Marchisio_R-SNN_IROS21} employs noise filtering to remove the adversarial attacks in the DVS inputs. The experiments demonstrate that such noise filtering slightly affects the SNN outputs for clean event sequences, while a wide range of filter parameters can increase the robustness of the SNN under attack by up to 90\%.

\section{Conclusion}

The use of Edge AI and tinyML systems is expected to grow fast in the coming years. 
Therefore, ensuring their high energy efficiency and robustness is important. 
This paper provides an overview of challenges and potential solutions for improving performance, energy efficiency, and robustness (i.e., reliability and security) of Edge AI. 
It shows that HW/SW co-design and co-optimization techniques at the design- and run-time can be combined through a cross-layer framework to efficiently address these challenges.

\section*{Acknowledgments}
This work was partly supported by Intel Corporation through Gift funding for the project ”Cost-Effective Dependability for Deep Neural Networks and Spiking Neural Networks”.

\bibliographystyle{IEEEtran}
\bibliography{bibliography}

\end{document}